\documentclass{ws-mpla}

\newcommand{\reseteqnum}{\setcounter{equation}{0}}
\newcommand{\nn}{\nonumber}
\newcommand{\eqn}[1]{(\ref{#1})}
\newcommand{\ovl}[1]{\overline{#1}}

\newcommand{\wh}[1]{\widehat{#1}}
\newcommand{\p}{\partial}
\newcommand{\dslash}{\p\kern-1.2ex /}
\newcommand{\Dslash}{D\kern-1.5ex /}
\newcommand{\bpsi}{{\overline{\psi}}}

\newcommand{\brho}{{\overline{\rho}}}
\newcommand{\bP}{\Pi}
\newcommand{\hP}{{\widehat{\Pi}}}
\newcommand{\oP}{{\overline{\Pi}}}
\newcommand{\ba}{{\overline{a}}}
\newcommand{\bq}{{\overline{q}}}
\newcommand{\tr}{{\rm tr}}

\newcommand{\braket}[2]{\vev{#1 | #2}}
\newcommand{\vev}[1]{\left\langle #1 \right\rangle}

\begin{document}

\markboth{Yusuke Taniguchi}
{Schr\"odinger functional formalism for overlap Dirac operator and
domain-wall fermion}

%
\catchline{}{}{}{}{}
%

\title{
%
%
Schr\"odinger functional formalism for\\
overlap Dirac operator and domain-wall fermion
\footnote{Conference report for ``International Conference on
Non-Perturbative Quantum Field Theory: Lattice and Beyond'',
Zhongshan University, Guangzhou, China, December 18-20, 2004.}
}

\author{\footnotesize Yusuke Taniguchi}

\address{
Institute of Physics, University of Tsukuba, Tsukuba, Ibaraki 305-8571,
Japan
\\
tanigchi@het.ph.tsukuba.ac.jp}

\maketitle

\pub{Received (Day Month Year)}{Revised (Day Month Year)}

\begin{abstract}
 In this proceeding we propose a new procedure to impose the Schr\"odinger
 functional Dirichlet boundary condition on the overlap Dirac operator
 and the domain-wall fermion using an orbifolding projection.
 With this procedure the zero mode problem with Dirichlet boundary
 condition can easily be avoided.

\end{abstract}

\ccode{PACS Nos.: include PACS Nos.}

\reseteqnum
\section{Introduction}

The Schr\"odinger functional (SF) is defined as a transition amplitude
between two boundary states with finite time separation
\cite{Symanzik,Luescher85} 
\begin{eqnarray}
Z=\braket{C';x_0=T}{C;x_0=0}=\int{\cal D}\Phi e^{-S[\Phi]}
\end{eqnarray}
and is written in a path integral representation of the field theory
with some specific boundary condition.
One of applications of the SF is to define a renormalization scheme
beyond perturbation theory, where the renormalization scale is given by
a finite volume $T\times L^3\sim L^4$ of the system.
The formulation is already accomplished for the non-linear
$\sigma$-model \cite{LWW}, the non-Abelian gauge theory \cite{LNWW} and
the QCD with the Wilson fermion \cite{Sint94,Sint95} including
${\cal O}(a)$ improvement procedure \cite{LSSW9605,LW96}.
(See Ref.~\cite{Sint01} for review.)

In this formalism several renormalization quantities like running gauge
coupling \cite{LSWW93,LSWW94,alpha95,LW95,NW,SS},
Z-factors and ${\cal O}(a)$ improvement factors
\cite{JLLSSSWW,LSSWW,LSSW9611,JS,CLSW}
are extracted conveniently by using a Dirichlet boundary
conditions for spatial component of the gauge field
\begin{eqnarray}
 A_k(x)|_{x_0=0}=C_k(\vec{x}),\quad
 A_k(x)|_{x_0=T}=C_k'(\vec{x})
\label{eqn:DBCgauge}
\end{eqnarray}
and for the quark fields
\begin{eqnarray}
&&
P_+\psi(x)|_{x_0=0}=\rho(\vec{x}),\quad
P_-\psi(x)|_{x_0=T}=\rho'(\vec{x}),
\label{eqn:DBCpsi}
\\&&
\bpsi(x)P_-|_{x_0=0}=\brho(\vec{x}),\quad
\bpsi(x)P_+|_{x_0=T}=\brho'(\vec{x}),
\label{eqn:DBCbpsi}
\\&&
P_\pm=\frac{1\pm\gamma_0}{2}.
\end{eqnarray}
One of privilege of this Dirichlet boundary condition is that the system
acquire a mass gap and there is no infra-red divergence.

We notice that the boundary condition is not free to set since it
generally breaks symmetry of the theory and may affect
renormalizability.
However the field theory with Dirichlet boundary condition is shown to
be renormalizable for the pure gauge theory \cite{LNWW}.
And it is also the case for the Wilson fermion \cite{Sint95} by
including a shift in the boundary fields.

Although it is essential to adopt Dirichlet boundary condition for a
mass gap and renormalizability, it has a potential problem of zero mode
in fermion system.
For instance starting from a free Lagrangian
\begin{eqnarray}
{\cal L}=\bpsi\left(\gamma_\mu\p_\mu+m\right)\psi
\end{eqnarray}
with positive mass $m>0$ and the Dirichlet boundary condition
\begin{eqnarray}
P_-\psi|_{x_0=0}=0,\quad
P_+\psi|_{x_0=T}=0
\end{eqnarray}
the zero eigenvalue equation $\left(\gamma_0\p_0+m\right)\psi=0$ in
temporal direction allows a solution
\begin{eqnarray}
\psi=P_+e^{-mx_0}+P_-e^{-m(T-x_0)}
\end{eqnarray}
in $T\to\infty$ limit and a similar solution remains even for finite $T$
with an exponentially small eigenvalue $\propto e^{-mT}$.
In the SF formalism this solution is forbidden by adopting an ``opposite''
Dirichlet boundary condition \eqn{eqn:DBCpsi} and the system has a
finite gap even for $m=0$ \cite{Sint94}.

In the SF formalism of the Wilson fermion \cite{Sint94} we cut the
Wilson Dirac operator at the boundary and the Dirichlet boundary
condition is automatically chosen among
\begin{eqnarray}
P_\pm\psi|_{x_0=0}=0,\quad
P_\mp\psi|_{x_0=T}=0
\end{eqnarray}
depending on signature of the Wilson term.
For example if we adopt a typical signature of the Wilson term
\begin{eqnarray}
D_W=\gamma_\mu\frac{1}{2}\left(\nabla_\mu^*+\nabla_\mu\right)
-\frac{a}{2}\nabla_\mu^*\nabla_\mu+M
\label{eqn:DW}
\end{eqnarray}
the allowed Dirichlet boundary condition is the same as
\eqn{eqn:DBCpsi}.
In this case the zero mode solution can be forbidden by choosing a
proper signature for the mass term; the mass should be kept positive
$M\ge0$ to eliminate the zero mode \cite{Sint94}.

However this zero mode problem may become fatal in the Ginsparg-Wilson
fermion including the overlap Dirac operator
\cite{Neuberger97,Neuberger98} and the domain-wall fermion
\cite{Kaplan,Shamir,FS,KN}.
The overlap Dirac operator is defined by using the Wilson Dirac operator
\eqn{eqn:DW} as
\begin{eqnarray}
D=\frac{1}{\ba}\left(1+D_W\frac{1}{\sqrt{D_W^\dagger D_W}}\right),\quad
\ba=\frac{a}{|M|}.
\label{eqn:OD}
\end{eqnarray}
Here we notice that the Wilson fermion mass $M$ should be kept negative
in a range $-2<M<0$ to impose heavy masses on the doubler modes and a
single massless mode to survive.
As explained in the above when the Dirichlet boundary condition is
imposed directly to the kernel $D_W$ an exponentially small eigenvalue
is allowed for this choice of the Wilson parameter and the mass.
Nearly zero eigenvalue in $D_W$ may break the locality of the overlap
Dirac operator \cite{HJL}.
The situation is also quite similar in the domain-wall fermion.
The same zero mode solution appears in the transfer matrix in fifth
direction, which suppresses the dumping solution in fifth dimension and
allow a chiral symmetry breaking term to appear in the Ward-Takahashi
identity \cite{FS}.

Since a naive formulation of the SF formalism by setting the Dirichlet
boundary condition for the kernel $D_W$ does not work, we need different
procedure to impose boundary condition on the overlap Dirac fermion.
In this paper we propose an orbifolding projection for this purpose.
In section \ref{sec:continuum} we introduce an orbifolding in a
continuum theory and show that the Dirac operator of the orbifolded
theory satisfy the same SF boundary condition.
As applications of this procedure we consider the overlap Dirac
fermion in section \ref{sec:overlap} and the domain-wall fermion in
section \ref{sec:dwf}.
Section \ref{sec:conclusion} is devoted for conclusion.

\section{Orbifolding for continuum theory}
\label{sec:continuum}

We notice a fact that the chiral symmetry is broken explicitly by the
Dirichlet boundary condition \eqn{eqn:DBCpsi} in the SF formalism.
This should be also true in the overlap Dirac operator; the
Ginsparg-Wilson relation should be broken in some sense, which was not
accomplished in a naive formulation.
We would adopt this property as a criterion of the SF formalism.

Then we remind a fact that an orbifolded field theory is
equivalent to a field theory with some specific boundary condition.
Since it is possible to break chiral symmetry by an orbifolding
projection in general, it may be able to represent the SF formalism
as an orbifolded theory.
In this section we search for an orbifolding projection which is not
consistent with chiral symmetry and provide the same SF Dirichlet
boundary condition \eqn{eqn:DBCpsi} and \eqn{eqn:DBCbpsi} at fixed
points.

We consider a massless free fermion on $S^1\times\bm{R}^3$
\begin{eqnarray}
{\cal L}=\bpsi(x)\gamma_\mu\p_\mu\psi(x),
\end{eqnarray}
where the anti-periodic boundary condition is set in temporal direction
of length $2T$
\begin{eqnarray}
\psi(\vec{x},x_0+2T)=-\psi(\vec{x},x_0),\quad
\bpsi(\vec{x},x_0+2T)=-\bpsi(\vec{x},x_0).
\label{eqn:APB}
\end{eqnarray}

The orbifolding $S^1/Z_2$ in temporal direction is accomplished by
identifying the negative time with the positive one
$x_0\leftrightarrow-x_0$.
Identification of the fermion field is given by using a symmetry
transformation including the time reflection
\begin{eqnarray}
\psi(x) \to \Sigma\psi(x),\quad
\bpsi(x) \to \bpsi(x)\Sigma,\quad
\Sigma=i\gamma_5\gamma_0R,
\label{eqn:timeref0}
\end{eqnarray}
where $R$ is a time reflection operator
\begin{eqnarray}
R\psi(\vec{x},x_0)=\psi(\vec{x},-x_0).
\end{eqnarray}
$R$ has two fixed points $x_0=0,T$, where $x_0=0$ is a symmetric and
$x_0=T$ is an anti-symmetric fixed point because of the anti-periodicity
\begin{eqnarray}
R\psi(\vec{x},0)=\psi(\vec{x},0),\quad
R\psi(\vec{x},T)=-\psi(\vec{x},T).
\end{eqnarray}
It is free to add any internal symmetry transformation for the
identification and we use the chiral symmetry of the massless fermion
\begin{eqnarray}
\psi(x)\to-i\gamma_5\psi(x),\quad
\bpsi(x)\to-\bpsi(x)i\gamma_5.
\label{eqn:chiraltr}
\end{eqnarray}
Combining \eqn{eqn:timeref0} and \eqn{eqn:chiraltr} we have the
orbifolding symmetry transformation
\begin{eqnarray}
\psi(x) \to -\Gamma\psi(x),\quad
\bpsi(x) \to \bpsi(x)\Gamma,\quad
\Gamma=\gamma_0R.
\label{eqn:orbifoldingtr}
\end{eqnarray}

The orbifolding of the fermion field is given by selecting the following
symmetric sub-space
\begin{eqnarray}
\bP_+\psi(x)=0,\quad
\bpsi(x)\bP_-=0,\quad
\bP_\pm=\frac{1\pm\Gamma}{2}.
\end{eqnarray}
We notice that this orbifolding projection provides the proper
homogeneous SF Dirichlet boundary condition at fixed points $x_0=0,T$
\begin{eqnarray}
&&
P_+\psi(x)|_{x_0=0}=0,\quad
P_-\psi(x)|_{x_0=T}=0,
\label{eqn:DBC1}
\\&&
\bpsi(x)P_-|_{x_0=0}=0,\quad
\bpsi(x)P_+|_{x_0=T}=0.
\label{eqn:DBC2}
\end{eqnarray}
The orbifolded action is given by the same projection
\begin{eqnarray}
S=\frac{1}{2}\int d^4x\bpsi(x)D_{\rm SF}\psi(x),\quad
D_{\rm SF}=\bP_+\dslash\bP_-,
\end{eqnarray}
where factor $1/2$ is included since the temporal direction is doubled
compared to the original SF formalism.
We notice that the chiral symmetry is broken explicitly for $D_{\rm SF}$
by the projection.

Now we have two comments.
Since the Schr\"odinger functional of the pure gauge theory is already
well established \cite{LNWW} we treat the gauge field as an external
field and adopt a configuration which is time reflection invariant
\begin{eqnarray}
A_0(\vec{x},-x_0)=-A_0(\vec{x},x_0),\quad
A_i(\vec{x},-x_0)=A_i(\vec{x},x_0)
\end{eqnarray}
and satisfy the SF boundary condition \eqn{eqn:DBCgauge}
simultaneously.
We set periodic boundary condition for the gauge field
\begin{eqnarray}
A_\mu(\vec{x},x_0+2T)=A_\mu(\vec{x},x_0).
\end{eqnarray}
Second comment is on the mass term.
Although the mass term is not consistent with the chiral symmetry,
we can find a symmetric mass term under the orbifolding transformation.
A requirement is that the mass matrix $M$ should anti-commute with
orbifolding operator $\left\{M,\Gamma\right\}=0$.
One of the candidate is a time dependent mass $M=m\eta(x_0)$ with
anti-symmetric and periodic step function
\begin{eqnarray}
&&
\eta(-x_0)=-\eta(x_0),\quad
\eta(x_0+2T)=\eta(x_0),
\nn\\&&
\eta(x_0)=1\quad{\rm for}\quad 0< x_0 <T.
\label{eqn:eta}
\end{eqnarray}

Now the Dirac operator becomes
\begin{eqnarray}
D(m)=\gamma_\mu\left(\p_\mu-iA_\mu(x)\right)+m\eta(x_0),
\end{eqnarray}
which has the orbifolding symmetry $\left\{D(m),\Gamma\right\}=0$.
The orbifolded Dirac operator is defined by the projection
\begin{eqnarray}
D_{\rm SF}(m)=\bP_+D(m)\bP_-.
\end{eqnarray}
Once the orbifold construction gives the Dirac operator with SF boundary
condition, the spectrum and the propagator are uniquely determined to be
equivalent to those of Ref.~\refcite{Sint94} and Ref.~\refcite{LW96}.
One can easily check that this is the case at tree level.

For example eigenvalues of the free SF Dirac operator $D_{\rm SF}$ is
derived as follows.
We first notice that the SF Dirac operator connects two different
Hilbert sub-space
\begin{eqnarray}
D_{\rm SF} : {\cal H}_+\to{\cal H}_-,\quad
D_{\rm SF}^\dagger : {\cal H}_-\to{\cal H}_+,
\end{eqnarray}
where
${\cal H}_\pm=\left\{\psi|\bP_\pm\psi=0\right\}$.
As in the original SF formulation \cite{Sint94} it is necessary to
introduce a ``doubled'' Hermitian Dirac operator
\begin{eqnarray}
{\cal D}=\pmatrix{&D_{\rm SF}\cr D_{\rm SF}^\dagger\cr},
\label{eqn:doubledD}
\end{eqnarray}
which connects the same Hilbert space
${\cal D}:{\cal H}_-\oplus{\cal H}_+\to{\cal H}_-\oplus{\cal H}_+$
in order to make the eigenvalue problem to be well defined.
This Dirac operator acts on a ``two component'' vector
\begin{eqnarray}
\Psi=\pmatrix{\psi_-\cr\psi_+\cr},\quad
\psi_-\in{\cal H}_-,\quad
\psi_+\in{\cal H}_+
\end{eqnarray}
and the eigenvalue problem is given in a following form
\begin{eqnarray}
D_{\rm SF}\psi_+=\lambda\psi_-,\quad
D_{\rm SF}^\dagger\psi_-=\lambda\psi_+
\label{eqn:ev}
\end{eqnarray}
with a real eigenvalue $\lambda$.
In the following we consider one dimensional problem for simplicity
\begin{eqnarray}
D_{\rm SF}(m)=\bP_+D(m)\bP_-,\quad
D(m)=\gamma_0\p_0+m\eta(x_0)
\end{eqnarray}
with an eigenvalue $\lambda_0$.

A candidate of the eigen-function $f_\pm\in{\cal H}_\pm$ satisfying
anti-periodicity in $2T$ is given by
\begin{eqnarray}
&&
f_+(x_0)=\alpha P_+S_-(x_0)+\beta P_-S_+(x_0),
\label{eqn:ef1}
\\&&
f_-(x_0)=\alpha P_+S_+(x_0)+\beta P_-S_-(x_0),
\label{eqn:ef2}
\end{eqnarray}
where $\alpha$, $\beta$ are normalization constant and $S_\pm$ are
defined for each bulk region with a cusp at the boundary
\begin{eqnarray}
&&
S_-(x_0)=(-)^n\sin p_0\left(x_0-2nT\right)\quad{\rm for}\quad
(2n-1)T\le x_0\le(2n+1)T,
\nn\\&&
S_+(x_0)=(-)^{n+1}\sin p_0\left(x_0-(2n+1)T\right)\quad{\rm for}\quad
2nT\le x_0\le2(n+1)T.
\nn
\end{eqnarray}
The eigenvalue equation \eqn{eqn:ev} has solution on these functions
only when a quantization condition is satisfied for $p_0$
\begin{eqnarray}
\tan p_0T=-\frac{p_0}{m}
\label{eqn:quantization}
\end{eqnarray}
and the eigenvalue becomes
\begin{eqnarray}
\lambda_0=-\frac{m}{\cos p_0T}=\frac{p_0}{\sin p_0T},\quad
\lambda_0^2=p_0^2+m^2.
\end{eqnarray}
This result agrees with that of Ref.\refcite{Sint94}.

When a spatial momentum is introduced we are to solve an eigenvalue
problem of the matrix \cite{Sint94}
\begin{eqnarray}
{\cal C}=\pmatrix{&C\cr C^\dagger\cr},\quad
C=i\gamma_kp_k+\lambda_0
\end{eqnarray}
and the eigenvalue $\lambda$ of the four dimensional Dirac operator is
given by
\begin{eqnarray}
\lambda^2=p_0^2+\vec{p}^2+m^2.
\label{eqn:eigenvalue}
\end{eqnarray}

\section{Orbifolding for overlap Dirac fermion}
\label{sec:overlap}

Application of the orbifolding procedure is straightforward to the
Ginsparg-Wilson fermions including the overlap Dirac operator
\cite{Neuberger97,Neuberger98}, the domain-wall fermion
\cite{Kaplan,Shamir,FS,KN} and the perfect action
\cite{DHHKN,BW,Hasenfratz} which possess both the time reflection
symmetry
\begin{eqnarray}
\left[\Sigma,D\right]=0
\label{eqn:timeref1}
\end{eqnarray}
and the lattice chiral symmetry \cite{Luescher98} stemming from
the Ginsparg-Wilson relation \cite{GW}
\begin{eqnarray}
&&
\gamma_5D+D\gamma_5=\ba D\gamma_5D.
\label{eqn:GW}
\end{eqnarray}
In this subsection we concentrate on the overlap Dirac operator
\eqn{eqn:OD}, for which the time reflection symmetry \eqn{eqn:timeref1}
comes from that of the Wilson Dirac operator
$\left[\Sigma,D_W\right]=0$.

\subsection{Orbifolding construction of Dirichlet boundary}

As in the continuum case we consider a massless fermion on a
lattice $2N_T\times N_L^3$ with anti-periodic boundary condition in temporal
direction \eqn{eqn:APB}.
We use an orbifolding $S^1/Z_2$ in temporal direction.
Identification of the fermion field is given by using the time
reflection \eqn{eqn:timeref0} and the chiral symmetry of the overlap
Dirac fermion \cite{Luescher98}
\begin{eqnarray}
\psi(x)\to-i\wh{\gamma}_5\psi(x),\quad
\bpsi(x)\to-\bpsi(x)i\gamma_5,\quad
\wh{\gamma}_5=\gamma_5\left(1-\ba D\right),
\label{eqn:chiraltr2}
\end{eqnarray}
where the gauge field is treated as an external field and we adopt a time
reflection symmetric configuration
\begin{eqnarray}
&&
U_k(\vec{x},x_0)=U_k(\vec{x},-x_0),\quad
U_0(\vec{x},x_0)=U_0^\dagger(\vec{x},-x_0-1),
\label{eqn:gauge1}
\end{eqnarray}
satisfying the SF Dirichlet boundary condition simultaneously
\begin{eqnarray}
&&
U_k(\vec{x},0)=W_k(\vec{x}),\quad
U_k(\vec{x},N_T)=W_k'(\vec{x}).
\label{eqn:gauge2}
\end{eqnarray}

Combining \eqn{eqn:timeref0} and \eqn{eqn:chiraltr2} we have the
orbifolding symmetry transformation
\begin{eqnarray}
\psi(x) \to -\wh{\Gamma}\psi(x),\quad
\bpsi(x) \to \bpsi(x)\Gamma,\quad
\wh{\Gamma}=\Gamma(1-\ba D),
\label{eqn:orbifoldingtr2}
\end{eqnarray}
where $\Gamma$ is the same as the continuum one \eqn{eqn:orbifoldingtr}.
We notice that starting from the time reflection symmetry of the Dirac
operator \eqn{eqn:timeref1} and the Ginsparg-Wilson relation
\eqn{eqn:GW} we have another GW relation for $\Gamma$
\begin{eqnarray}
\Gamma D+D\Gamma=\ba D\Gamma D
\end{eqnarray}
and $\Gamma$ Hermiticity
\begin{eqnarray}
\Gamma D\Gamma=D^\dagger.
\end{eqnarray}
The operator $\wh{\Gamma}$ has a property $\wh{\Gamma}^2=1$ like
$\Gamma$ and can be used to define a projection operator in the
following.

The orbifolding identification of the fermion field is given in the same
way with slightly different projection operator
\begin{eqnarray}
\hP_+\psi(x)=0,\quad
\bpsi(x)\bP_-=0,\quad
\hP_\pm=\frac{1\pm\wh{\Gamma}}{2},
\label{eqn:orbifolding}
\end{eqnarray}
which turn out to be the SF Dirichlet boundary condition \eqn{eqn:DBC1}
and \eqn{eqn:DBC2} at fixed points in the continuum limit.
Using the time reflection symmetry \eqn{eqn:timeref1} we can easily show
that the projection operators $\Gamma$ and $\wh{\Gamma}$ do not have an
``index''
\begin{eqnarray}
\tr\Gamma=\tr\hat{\Gamma}=0
\end{eqnarray}
and furthermore we can find a local unitary transformation
\begin{eqnarray}
u=\frac{1+\Sigma}{2}(1-\ovl{a}D)+\frac{1-\Sigma}{2},\quad
u'=\gamma_5u\gamma_5,
\label{eqn:unitary}
\end{eqnarray}
which connects $\wh{\Gamma}$ and $\Gamma$ as
\begin{eqnarray}
\wh{\Gamma}=u^\dagger\Gamma u,\quad
\wh{\Gamma}=u'\Gamma u'^\dagger.
\end{eqnarray}
The projection operator $\hP_\pm$ spans essentially the same Hilbert
sub-space as $\bP_\pm$.
We notice that this unitary operator connects $\wh{\gamma}_5$ and
$\gamma_5$ in a similar way
\begin{eqnarray}
\wh{\gamma}_5=u^\dagger\gamma_5 u,\quad
\wh{\gamma}_5=u'\gamma_5 u'^\dagger.
\end{eqnarray}

The physical quark operator is defined to transform in a same manner as
the continuum under chiral rotation,
\begin{eqnarray}
\delta q(x)=\gamma_5q(x),\quad
\delta\bq(x)=\bq(x)\gamma_5.
\end{eqnarray}
Since we have a unitary operator $u$ and $u'$ we have several ways 
to define a physical quark field from GW fermion fields $\psi$ and
$\bpsi$.
For example
\begin{eqnarray}
&&
q(x)=\left(1-\frac{\ba}{2}D\right)\psi(x),\quad
\bq(x)=\bpsi(x),
\label{eqn:quark1}
\\&&
q(x)=u\psi(x),\quad
\bq(x)=\bpsi(x),
\label{eqn:quark2}
\\&&
q(x)=u'^\dagger\psi(x),\quad
\bq(x)=\bpsi(x).
\label{eqn:quark3}
\end{eqnarray}
These three definitions are not independent but connected with
$u+u'^\dagger=\left(2-\ba D\right)$.
The orbifolding of the physical quark field becomes the same as that of
the continuum theory
\begin{eqnarray}
\bP_+q(x)=0,\quad
\bq(x)\bP_-=0.
\end{eqnarray}

The massless orbifolded action is given by
\begin{eqnarray}
S=\frac{1}{2}a^4\sum\bpsi D_{\rm SF}\psi,\quad
D_{\rm SF}=\bP_+D\hP_-.
\label{eqn:massless}
\end{eqnarray}
We have four comments here.
(i) It should be emphasized that the SF Dirac operator
$D_{\rm SF}$ is local since it is constructed by multiplying local
objects only.
(ii) The massless SF Dirac operator $D_{\rm SF}$ does not satisfy the
chiral Ginsparg-Wilson relation \eqn{eqn:GW}.
(iii) Although two different projection operators $\Gamma$ and
$\wh{\Gamma}$ are used from the left and right of $D_{\rm SF}$ this does
not bring the problem we encountered in the chiral gauge theory since
these two operators are connected by the unitary transformation $u$ or
$u'$.

\subsection{Surface term}

When extracting the renormalization factors of fermions it is
convenient to consider a operator involving the boundary source fields
\begin{eqnarray}
&&
\zeta(\vec{x})=\frac{\delta}{\delta\ovl{\rho}(\vec{x})},\quad
\ovl{\zeta}(\vec{x})=-\frac{\delta}{\delta\rho(\vec{x})},
\\&&
\zeta'(\vec{x})=\frac{\delta}{\delta\ovl{\rho}'(\vec{x})},\quad
\ovl{\zeta}'(\vec{x})=-\frac{\delta}{\delta\rho'(\vec{x})},
\end{eqnarray}
where $\rho, \cdots, \ovl{\rho}'$ are boundary values of the fermion
fields given in \eqn{eqn:DBCpsi} and \eqn{eqn:DBCbpsi}.
Coupling of the boundary value to the bulk dynamical fields was
naturally introduced in the Wilson fermion \cite{Sint94}.
However this is not the case for our construction since the boundary
value vanishes with the orbifolding projection.

In this paper we regard the boundary value as an external source field
and introduce its coupling with the bulk fields according to the
criteria: the coupling terms (surface terms) are local and reproduce the
same form of the correlation function between the boundary fields in the
continuum limit.
Here we define the boundary vales on the physical quark fields
\begin{eqnarray}
&&
P_+q(x)|_{x_0=0}=\rho(\vec{x}),\quad
P_-q(x)|_{x_0=N_T}=\rho'(\vec{x}),
\label{eqn:DBCq}
\\&&
\bq(x)P_-|_{x_0=0}=\brho(\vec{x}),\quad
\bq(x)P_+|_{x_0=N_T}=\brho'(\vec{x}).
\label{eqn:DBCbq}
\end{eqnarray}
One of candidates of the surface term is
\begin{eqnarray}
S_{\rm surface}&=&a^3\sum_{\vec{x}}\Bigl(
-\left.\ovl{\rho}(\vec{x})P_-q(x)\right|_{x_0=0}
-\left.\bq(x)P_+\rho(\vec{x})\right|_{x_0=0}
\nn\\&&
-\left.\ovl{\rho}'(\vec{x})P_+q(x)\right|_{x_0=N_T}
-\left.\bq(x)P_-\rho'(\vec{x})\right|_{x_0=N_T}
\Bigr),
\label{eqn:surface}
\end{eqnarray}
where $q$ and $\bq$ are active dynamical fields on the boundary.

According to Ref.~\refcite{LW96} we introduce the generating functional
\begin{eqnarray}
Z_F\left[\ovl{\rho}',\rho';\ovl{\rho},\rho;\ovl{\eta},\eta;U\right]&=&
\int D\psi D\bpsi \exp\biggl\{
-S_F\left[U,\bpsi,\psi;\ovl{\rho}',\rho',\ovl{\rho},\rho\right]
\nn\\&&
+a^4\sum_{x}\left(\bpsi(x)\eta(x)+\ovl{\eta}(x)\psi(x)\right)
\biggr\},
\end{eqnarray}
where $\eta(x)$ and $\ovl{\eta}(x)$ are source fields for the fermion
fields and the total action $S_F$ is given as a sum of the bulk action
\eqn{eqn:massless} and the surface term \eqn{eqn:surface}.
We notice that the fermion fields $\psi$ and $\bpsi$ obey the
orbifolding condition \eqn{eqn:orbifolding}.
The correlation functions between the boundary fields are derived
according to ordinary procedures of perturbation theory.

\subsection{Phase of Dirac determinant}

In general the determinant of the SF Dirac operator is not real for the
overlap fermion since there is no $\gamma_5$ Hermiticity.
Instead we have a following ``Hermiticity'' relation
\begin{eqnarray}
\gamma_5uD_{\rm SF}u^\dagger\gamma_5=D_{\rm SF}^\dagger,\quad
\gamma_5u'^\dagger D_{\rm SF}u'\gamma_5=D_{\rm SF}^\dagger.
\label{eqn:hermiticity}
\end{eqnarray}
However one cannot conclude reality from this relation since the SF
Dirac operator connects different Hilbert sub-space as
\begin{eqnarray}
&&
D_{\rm SF} : \wh{\cal H}_+\to{\cal H}_-,\quad
\wh{\cal H}_+=\left\{\psi|\hP_+\psi=0\right\},\quad
{\cal H}_-=\left\{\psi|\Pi_-\psi=0\right\}.
\label{eqn:subspace}
\end{eqnarray}
and the determinant cannot be evaluated directly with $D_{\rm SF}$.
We need to make a ``Hermitian'' Dirac operator which connects the
same Hilbert sub-space in order to define the Dirac determinant.
This is accomplished by $u^\dagger\gamma_5$ or $u'\gamma_5$, which
turns out to be $\gamma_5$ in the continuum.
We define
\begin{eqnarray}
&&
H_{\rm SF}=D_{\rm SF}u^\dagger\gamma_5=\bP_+Du^\dagger\gamma_5\bP_+
\quad:\quad{\cal H}_-\to{\cal H}_-,
\\&&
H_{\rm SF}'=D_{\rm SF}u'\gamma_5=\bP_+Du'\gamma_5\bP_+
\quad:\quad{\cal H}_-\to{\cal H}_-.
\end{eqnarray}
The determinant is evaluated on the sub-space ${\cal H}_-$
\begin{eqnarray}
&&
\det_{\{{\cal H_-}\}}H_{\rm SF}
=\det\left(\bP_+Du^\dagger\gamma_5\bP_++\bP_-\right),
\\&&
\det_{\{{\cal H_-}\}}H_{\rm SF}'
=\det\left(\bP_+Du'\gamma_5\bP_++\bP_-\right),
\end{eqnarray}
where the right hand side is understood to be evaluated in the full
Hilbert space by filling the opposite sub-space ${\cal H}_+$ with
unity.

The phase of the determinant is given as follows
\begin{eqnarray}
&&
\left(\det_{\{{\cal H_-}\}}H_{\rm SF}^{(\prime)}\right)^*=
e^{-2i\phi^{(\prime)}}\left(\det_{\{{\cal H_-}\}}H_{\rm SF}^{(\prime)}\right),
\\&&
e^{-2i\phi}=\det_{\{{\cal H_-}\}}\left(\gamma_5u\right)^2=\det u,
\\&&
e^{-2i\phi'}=\det_{\{{\cal H_-}\}}\left(\gamma_5u'^\dagger\right)^2
=\det u^\dagger=e^{2i\phi},
\end{eqnarray}
which is not real in general.
The determinant of the unitary operator $u$ is given by a product of
eigenvalues $\lambda_n$ of the overlap Dirac operator 
\begin{eqnarray}
\det u=\prod_{n\in\{+\}}(1-a\lambda_n),
\end{eqnarray}
where product is taken over a sub-space in which the eigenvalue of
$\Sigma=+1$ and the conjugate eigenvalue $\lambda_n^*$ does not
necessarily belongs to this sub-space.

However we notice that this complexity of the Dirac determinant is not an
essential problem since the phase is an ${\cal O}(a)$ irrelevant effect
and disappears in the continuum limit.
Furthermore if we consider variation of the phase
\begin{eqnarray}
\delta_{\epsilon(x)}\phi=\frac{i}{2}\tr\delta_{\epsilon(x)}uu^{-1}
=-\frac{i}{4}a\tr\left[\Sigma\delta_{\epsilon(x)}D\left(1-aD^\dagger\right)
\right]
\label{eqn:phasevar}
\end{eqnarray}
under a local variation of the link variable
\begin{eqnarray}
\delta_{\epsilon(x)}U_\mu(x)=a\epsilon_\mu(x)U_\mu(x)
\end{eqnarray}
we can show that $\delta_{\epsilon(x)}\phi$ is localized at the
boundary.
Since $\Sigma$ contains time reflection $R$ and both of the operator
$\delta D$ and $\left(1-aD^\dagger\right)$ are local, the trace in
\eqn{eqn:phasevar} has a contribution only at the boundary.
Contribution from the bulk is suppressed exponentially by the locality
property.

\subsection{Mass term}

The mass term may be introduced with the same procedure as the continuum
theory.
We consider a mass matrix $M$ which is consistent with the orbifolding
symmetry
\begin{eqnarray}
\Gamma M+M\wh{\Gamma}=0.
\label{eqn:symmass}
\end{eqnarray}
Since the orbifolding transformation is the same as the continuum one on
the physical quark fields, a naive candidate is to couple the continuum
mass matrix $m\eta(x_0)$ to the physical scalar density consisting of
$q(x)$ and $\bq(x)$.
Corresponding to various definition of the quark fields
\eqn{eqn:quark1}-\eqn{eqn:quark3} we have several definitions of the
mass term
\begin{eqnarray}
{\cal L}_m=
m\bpsi\eta\left(1-\frac{\ovl{a}}{2}D\right)\psi,\quad
m\bpsi\eta u\psi,\quad
m\bpsi\eta u'^\dagger\psi,
\end{eqnarray}
where $\eta$ is an anti-symmetric step function \eqn{eqn:eta} on
lattice.

However we encounter a problem with this naive definition of mass term,
since the massive Dirac operator does not satisfy the ``Hermiticity''
relation \eqn{eqn:hermiticity}.
The phase of the Dirac determinant becomes mass dependent although it is
still irrelevant ${\cal O}(a)$ term.
In order to avoid this unpleasant situation we may need even numbers of
flavors.

For two flavors case we define the two by tow Dirac operator as
\begin{eqnarray}
D_{\rm SF}^{(2)}(m)=\pmatrix{D_{\rm SF}(m)_1\cr&D_{\rm SF}(m)_2},
\end{eqnarray}
where
\begin{eqnarray}
&&
D_{\rm SF}(m)_1=\bP_+
\left(D+m\eta\left(1-\frac{\ovl{a}}{2}D\right)\right)\hP_-
\\&&
D_{\rm SF}(m)_2
=\bP_+\left(D+m\left(1-\frac{\ovl{a}}{2}D\right)u'\eta u'^\dagger\right)
\hP_-.
\end{eqnarray}
A ``Hermitian'' relation can be found for this two flavors Dirac
operator as
\begin{eqnarray}
D_{\rm SF}^{(2)}(m)^\dagger=
\tau^1\gamma_5UD_{\rm SF}^{(2)}(m)U^\dagger\gamma_5\tau^1,
\label{eqn:hermiticity2}
\end{eqnarray}
where $\tau^1$ and $U$ are two by two matrix acting on the flavor space
\begin{eqnarray}
\tau^1=\pmatrix{&1\cr1},\quad
U=\pmatrix{u\cr&u'^\dagger\cr}.
\end{eqnarray}
The Hermitian Dirac operator can be defined to connect the same Hilbert
sub-space as
\begin{eqnarray}
H_{\rm SF}^{(2)}(m)=D_{\rm SF}^{(2)}(m)U^\dagger\gamma_5\tau^1
\quad:\quad{\cal H}_-\oplus{\cal H}_-\to{\cal H}_-\oplus{\cal H}_-,
\end{eqnarray}
which is re-written in a trivially Hermitian form by a unitary matrix
$V$
\begin{eqnarray}
{H}_{\rm SF}^{(2)}(m)=
V\pmatrix{&D_{\rm SF}(m)_1\cr D_{\rm SF}(m)_1^\dagger\cr}V^\dagger,\quad
V=\pmatrix{1\cr&\gamma_5u}.
\end{eqnarray}
The determinant of this Dirac operator is evaluated in a single Hilbert
sub-space
\[\det_{\{{\cal H}_-\oplus{\cal H}_-\}}{H}_{\rm SF}\]
and becomes real.

\section{Orbifolding for domain-wall fermion}
\label{sec:dwf}

We consider the Shamir's domain-wall fermion \cite{Shamir,FS} on a
lattice $2N_T\times N_L^3\times N_5$ with anti-periodic boundary
condition in temporal direction
\begin{eqnarray}
&&
S=\sum_{\vec{x},\vec{y}}\sum_{x_0,y_0=-N_T+1}^{N_T}\sum_{s,t=1}^{N_5}
\bpsi(x,s)D_{\rm dwf}(x,y;s,t)\psi(y,t).
\end{eqnarray}
We adopt a notation used in CP-PACS collaboration and the Dirac operator
is given as a five dimensional Wilson's one with conventional Wilson
parameter $r=1$ and negative mass parameter $-M$ with $0<M<2$
\begin{eqnarray}
D_{\rm dwf}(x,y;s,t)&=&
\left(
 \frac{-1+\gamma_\mu}{2}U_\mu(x)\delta_{y_\mu,x_\mu+1}
+\frac{-1-\gamma_\mu}{2}U_\mu^\dagger(y)\delta_{y_\mu,x_\mu-1}\right)
\delta_{x_0,y_0}\delta_{s,t}
\nn\\&+&
\left(
\frac{-1+\gamma_5}{2}\Omega^+_{s,t}+\frac{-1-\gamma_5}{2}\Omega^-_{s,t}\right)
\delta_{x,y}
+(5-M)\delta_{x,y}\delta_{s,t},
\end{eqnarray}
where $\Omega^\pm$ are hopping operator in fifth direction with Dirichlet
boundary condition, whose explicit form is given by
\begin{eqnarray}
\Omega^+_{s,t}=\delta_{t,s+1},\quad
\Omega^+_{N_5,t}=0,\quad
\Omega^-=\left(\Omega^+\right)^\dagger.
\label{eqn:omega}
\end{eqnarray}
The physical quark field is defined by the fifth dimensional boundary
field with chiral projection
\begin{eqnarray}
&&
q(x)=\left(P_L\delta_{s,1}+P_R\delta_{s,N_5}\right)\psi(x,s),
\\&&
\bq(x)=\bpsi(x,s)\left(\delta_{s,N_5}P_L+\delta_{s,1}P_R\right),
\\&&
P_{R/L}=\frac{1\pm\gamma_5}{2}.
\end{eqnarray}
As in the formulation with the overlap Dirac operator the gauge
field is treated as an external field.

In order to apply the orbifolding construction of the SF Dirac operator
we need two symmetries of time reflection and chiral transformation.
The time reversal symmetry of the domain-wall fermion is given by
\begin{eqnarray}
&&
\psi(\vec{x},x_0,s)\to\ovl{\Sigma}_{x_0,y_0;s,t}\psi(\vec{x},y_0,t),
\quad
\bpsi(\vec{x},x_0,s)\to\bpsi(\vec{x},y_0,t)\ovl{\Sigma}_{y_0,x_0;t,s},
\label{eqn:timeref}
\\&&
\ovl{\Sigma}_{x_0,y_0;s,t}=i\gamma_5\gamma_0R_{x_0,y_0}P_{s,t},
\end{eqnarray}
where $P$ is a parity transformation in fifth direction
\begin{eqnarray}
P_{s,t}\psi(\vec{x},x_0,t)=\psi(\vec{x},x_0,N_5+1-s)
\end{eqnarray}
and $R$ is a time reflection operator acting on the temporal direction.
The domain-wall fermion Dirac operator is invariant under the time
reflection
\begin{eqnarray}
\left[\ovl{\Sigma},D_{\rm dfw}\right]=0
\label{eqn:timeref2}
\end{eqnarray}
since the reflection invariant gauge configuration \eqn{eqn:gauge1} and
\eqn{eqn:gauge2} is adopted .

The chiral transformation is given according to Ref.~\refcite{FS} by
rotating the fermion field vector like but with a different charge for
two boundaries in fifth direction
\begin{eqnarray}
\psi(x,s)\to iQ_{s,t}\psi(x,t),\quad
\bpsi(x,s)\to -\bpsi(x,t)iQ_{t,s},
\label{eqn:chiraltr3}
\end{eqnarray}
where $Q$ is an $s$ dependent charge
\begin{eqnarray}
Q_{s,t}={\rm Sgn}(N_5-2s+1).
\end{eqnarray}
Here we should notice that this chiral rotation is not an exact symmetry
of the domain-wall fermion Dirac operator but we have an explicit
breaking term
\begin{eqnarray}
QD_{\rm dwf}Q-D_{\rm dwf}=2X,
\end{eqnarray}
where $X$ is a contribution from the middle layer and picks up a charge
difference there
\begin{eqnarray}
X=\left(P_L\delta_{s,\frac{N_5}{2}}\delta_{t,\frac{N_5}{2}+1}
+P_R\delta_{s,\frac{N_5}{2}+1}\delta_{t,\frac{N_5}{2}}\right)\delta_{x,y}.
\end{eqnarray}
However it was discussed in Ref.~\refcite{FS} that if we consider the
correlation functions between the bilinear $\bpsi X\psi$ and the physical
quark operators the contribution is suppressed exponentially in $N_5$
under the condition that the transfer matrix in fifth direction has a
gap from unity.
Furthermore the domain-wall fermion with explicitly time reflection
invariant Dirac operator \eqn{eqn:timeref2} does not have index, since
the contribution to the index
\begin{eqnarray}
\lim_{N_5\to\infty}a^4\sum_{x}\vev{\bpsi(x,s)\gamma_5X_{s,t}\psi(x,t)}
=-\lim_{N_5\to\infty}\tr\left(\gamma_5X\frac{1}{D_{\rm dwf}}\right)
\end{eqnarray}
can be shown to vanish by using anti-commutativity
$\left\{\gamma_5X,\ovl{\Sigma}\right\}=0$.
We can expect no effect from $X$ to the physical theory and we shall
ignore this term in the following by constraining that we consider the
physical quark Green's functions only.

Combining \eqn{eqn:timeref} and \eqn{eqn:chiraltr3} we have the
orbifolding symmetry transformation
\begin{eqnarray}
&&
\psi(\vec{x},x_0,s)\to A_{x_0,y_0;s,t}\psi(\vec{x},y_0,t),\quad
\bpsi(\vec{x},x_0,s)\to\bpsi(\vec{x},y_0,t)A_{y_0,x_0;t,s},
\\&&
A_{x_0,y_0;s,t}=\gamma_0\gamma_5(PQ)_{s,t}R_{x_0,y_0},
\end{eqnarray}
where we used a relation
\begin{eqnarray}
\left\{P,Q\right\}=0.
\end{eqnarray}
The operator $A$ satisfy a property $A^2=1$ and can be used to define a
projection operator.
The orbifolding identification of the fermion field is given by
projecting onto the following symmetric sub-space
\begin{eqnarray}
\oP_-\psi(x,s)=0,\quad
\bpsi(x,s)\oP_-=0,\quad
\oP_\pm=\frac{1\pm{A}}{2}.
\end{eqnarray}
The orbifolding projection for the physical quark field is given by
picking up the boundary components from the projected fermion field,
which turns out to be the same condition as the continuum field.
The proper homogeneous SF Dirichlet boundary condition is provided at
fixed points.
The massless orbifolded action is given by
\begin{eqnarray}
S=\frac{1}{2}a^4\sum\bpsi D_{\rm SF}^{\rm dwf}\psi,\quad
D_{\rm SF}^{\rm dwf}=\oP_+D_{\rm dwf}\oP_+.
\end{eqnarray}
We have four comments here.
(i) We regard the orbifolding transformation as an exact symmetry of the
system by ignoring the explicit breaking term
\begin{eqnarray}
\left[A,D_{\rm dwf}\right]=0.
\label{eqn:symmetry}
\end{eqnarray}
(ii) The massless SF Dirac operator $D_{\rm SF}^{\rm dwf}$ breaks
``chiral symmetry'' under \eqn{eqn:chiraltr3} explicitly by the
projection $\oP_+$.
(iii) We have a Hermiticity relation for this Dirac operator
\begin{eqnarray}
\left(D_{\rm SF}^{\rm dwf}\right)^\dagger=
\gamma_5PD_{\rm SF}^{\rm dwf}\gamma_5P.
\end{eqnarray}
(iv) This Dirac operator connects the same Hilbert sub-space
\begin{eqnarray}
D_{\rm SF}^{\rm dwf} : \ovl{\cal H}_-\to\ovl{\cal H}_-,\quad
\ovl{\cal H}_-=\left\{\psi|\oP_-\psi=0\right\}.
\end{eqnarray}

\section{Conclusion}
\label{sec:conclusion}

In this paper we propose a new procedure to introduce the SF Dirichlet
boundary condition for general fermion fields.
Instead of cutting the Dirac operator at the boundary
we focus on a fact that the chiral
symmetry is broken explicitly in the SF formalism by the boundary
condition and adopt it as a criterion of the procedure.
We also notice that an orbifolded field theory is equivalent to a field
theory with some specific boundary condition.
We search for the orbifolding symmetry which is not consistent with the
chiral symmetry and reproduces the SF Dirichlet boundary condition on
the fixed points.
We found that the orbifolding $S^1/Z_2$ in temporal direction including
the time reflection, the chiral rotation and the anti-periodicity serves
this purpose well.

Application of this procedure to the overlap Dirac operator is
straightforward since this system has both the time reflection and the
chiral symmetry.
We found a technical problem that the Dirac determinant is complex.
However this is not essential since the phase of the determinant is an
irrelevant ${\cal O}(a)$ term.
The mass term may still have a problem.
We did not find a massive Dirac operator which has the same phase as the
massless one for a single flavor case.
For two flavors we can construct a Hermitian massive Dirac operator,
where the phase is absorbed into the fermion field.
However we should notice that the flavor symmetry is broken in this two
flavors formulation.
The SF formalism with the domain-wall fermion can be formulated in the
same way since the symmetry is exactly the same as the overlap
Dirac operator \cite{KN}.

\section*{Acknowledgement}

I would like to thank M.~L\"uscher for his valuable suggestions and
discussions.
Without his suggestions this work would not have completed.
I also thank to R.~Sommer, S.~Aoki, O.~B\"ar, T.~Izubuchi, Y.~Kikukawa
and Y.~Kuramashi for valuable discussions.

\end{document}